\documentclass[twocolumn,aps,prb,showpacs,amsmath,superscriptaddress,longbibliography,notitlepage]{revtex4-2}
\usepackage{amssymb}
\usepackage{mathrsfs}
\usepackage{graphicx}
\usepackage{float}
\usepackage[caption=false]{subfig}
\usepackage[pdftex,colorlinks,linkcolor={red!75!black},citecolor={red!75!black},urlcolor={red!75!black}]{hyperref}
\usepackage{verbatim}
\usepackage{txfonts}
\usepackage{epstopdf}
\usepackage[normalem]{ulem}
\usepackage{xcolor}
\usepackage{bm}
\usepackage[utf8]{inputenc}

\newcommand{\clr}{\color{red!75!black}}

\def\ket#1{\vert #1 \rangle}
\def\bra#1{\langle #1 \vert}

\begin{document}

\title{Non-Hermitian sensing via end-to-end Green's functions}

\author{Hui-Qiang Liang}
\affiliation{Department of Physics, Shandong University, Jinan 250100, China}
\author{Yu-Shi Cheng}
\affiliation{Department of Physics, Shandong University, Jinan 250100, China}
\author{Guo-Fu Xu}\email{xgf@sdu.edu.cn}
\affiliation{Department of Physics, Shandong University, Jinan 250100, China}

\date{\today}

\begin{abstract}
Sensing hinges on two key attributes: high sensitivity and resilience to noise. Here, we present a scheme that unites these features within a single framework. By harnessing the exponential amplification of end-to-end Green's functions—a direct signature of the non-Hermitian skin effect—our scheme achieves an exponential sensitivity scaling of $~e^{\alpha L}$ with $\alpha$ a model-specific constant and $L$ the system size. Notably, this enhanced sensitivity is protected by spectral winding numbers and endures under disorder when the non-Hermitian topological phase remains intact. We also explore its feasibility in synthetic non-Hermitian platforms, providing a pathway toward highly sensitive and noise-tolerant sensing.
\end{abstract}

\maketitle


\section*{Introduction}
Sensing serves as the interface between the quantum realm and measurable quantities like time, magnetic fields, and temperature~\citep{giovannetti2011advances,xiang2013quantum,toth2014quantum,szczykulska2016multi,degen_quantum_2017,pezze2018quantum,pirandola2018advances,aslam2023quantum}. Achieving both high sensitivity and robustness against imperfections is a central goal in the development of advanced sensing technologies. Among the emerging approaches in this direction, non-Hermitian sensing has attracted considerable attention~\citep{wiersig2020review,de2022non,wang2025sensing}. Non-Hermitian Hamiltonians are well known for providing an effective description of numerous non-conservative systems. Unlike their Hermitian counterparts, these systems exhibit a range of intrinsic properties arising from non-Hermiticity. Prominent examples include the non-Hermitian skin effect (NHSE)~\citep{lee2016anomalous,yao2018edge,yokomizo2019non,lee2019hybrid,luo2019higher,lee2019anatomy,liu2019second,okuma2020topological,li2020topological,borgnia2020non,zhang2020correspondence,zhang2022universal,li2022gain,li2023enhancement,nakamura_universal_2023,okuma2023non,lin2023topologically,zhu_brief_2023,gohsrich2025non}, characterized by the boundary localization of a large number of eigenstates, and the emergence of exceptional points (EPs)~\citep{kato1966perturbation,heiss1998collectivity,berry2004physics,rotter2009non,heiss2012physics,xu2017weyl,shen2018topological,ashida2020non,delplace2021symmetry,bergholtz2021exceptional}, where both eigenvalues and their associated eigenstates simultaneously coalesce. To date, numerous research efforts have exploited the exotic phenomena inherent to non-Hermitian systems to achieve enhanced sensitivity~\citep{wiersig_enhancing_2014,wiersig_sensors_2016,ren_ultrasensitive_2017,chen2017exceptional,sunada_large_2017,el-ganainy_non-hermitian_2018,budich_non-hermitian_2020,mcdonald_exponentially-enhanced_2020,bergholtz_exceptional_2021,koch_quantum_2022,luo_quantum_2022,yuan2023non,sarkar2024critical,rafi2024saturation,zhang2024true}, with some of these proposals already validated experimentally~\citep{liu_metrology_2016,chen_exceptional_2017,hodaei_enhanced_2017,lai_observation_2019,wang_petermann-factor_2020,yu_experimental_2020,kononchuk_exceptional-point-based_2022,li2023exceptional,deng2024ultrasensitive,parto2025enhanced,zhou2025nonhermitian}.

Green's functions constitute a cornerstone of modern theoretical physics, providing a powerful framework for describing the response of a physical system to external perturbations~\citep{economou2006green}. They have proven indispensable for computing transport properties, density of states, and correlation functions. Physically, a Green's function represents the propagator of a disturbance; it quantifies how a localized input at one point in space and time affects the system at another point. Mathematically, it is defined as the inverse of the differential operator that governs the equation of motion, and its poles directly encode the excitation spectrum of the system. In Hermitian systems, such responses generally decay with distance, which naturally precludes directional amplification. In non-Hermitian systems, however, this paradigm can be fundamentally altered~\citep{wanjura2020topological,helbig2020generalized,zirnstein2021bulk,wanjura2021correspondence,xue_simple_2021,li_quantized_2021,zirnstein2021exponentially,liang_anomalous_2022,hu2023green}. In particular, the NHSE can induce exponential amplification of the end-to-end Green's functions~\citep{li2022exact}. This observation naturally inspires us to develop a non-Hermitian sensing paradigm that leverages the end-to-end Green's function as a direct readout mechanism in non-Hermitian systems.

In this paper, we propose a non-Hermitian sensing scheme leveraging the end-to-end Green's functions of non-Hermitian systems exhibiting the NHSE. For a high-performance sensing scheme, two key features are indispensable: high sensitivity and resilience to noise. We show that our scheme unites these two features within a single framework. Specifically, Our scheme capitalizes on the exponential growth of end-to-end Green's functions, which arises as a direct manifestation of NHSE. This yields a sensitivity that scales as 
$~e^{\alpha L}$, where $\alpha$ is a model-specific constant and $L$ denotes the system size. Importantly, the enhanced sensitivity is safeguarded by spectral winding numbers, and it persists even in the presence of disorder, provided the non-Hermitian topological phase is not destroyed. We further examine the practical implementation of this approach within synthetic non-Hermitian systems, charting a viable route toward sensing with both ultrahigh sensitivity and noise resilience.

\section*{The basic idea} 
\label{SecII}
We consider a non-Hermitian lattice system that exhibits the non-Hermitian skin effect, with the corresponding Hamiltonian denoted by $H^0$. With the system configured as a sensor and the measurand included, the total Hamiltonian can be written as
\begin{equation}
\mathcal{H}=H^0+H^\Delta,
\end{equation}
where $\Delta$ represents the measurand, and $H^\Delta$, which is a function of $\Delta$, origins from the measurand interferes and can be regarded as a perturbation to the system. The Green's function $G_\text{out,in}$ corresponding to $\mathcal{H}=H^0+H^\Delta$ reads
\begin{equation}
G_\text{out,in} = \left[\frac{1}{\omega-\mathcal{H}}\right]_\text{out,in},
\end{equation}
where $\omega$ is the frequency of the input signal, and ``in" and ``out" are the lattice positions of the signal input and output respectively. In the sensing setup, the measurement signal reads
\begin{equation}
\mathcal{S}=\frac{\partial \big| G_\text{out,in} \big|}{\partial \Delta} \approx \frac{\big|G_\text{out,in}-G^0_\text{out,in}\big|}{\Delta}.
\end{equation} 
Here, we characterize the sensing performance by the response sensitivity $\mathcal{S}$ of the end-to-end Green’s function to the measurand $\Delta$.
In the above equation, the fact that $\Delta$ is a perturbation has been used and 
\begin{equation}
G^0_\text{out,in} = \left[\frac{1}{\omega-H^0}\right]_\text{out,in}
\end{equation}
is the Green's function corresponding to $H^0$. By using the perturbation theory of the Green's function~\citep{economou2006green}, we further obtain 
\begin{equation}
G_\text{out,in}-G^0_\text{out,in} = \left[[1-G^0H^\Delta]^{-1}G^0\right]_\text{out,in} - G^0_\text{out,in}.
\end{equation}
Since the perturbation $H^\Delta$ is weak, we expand the term $[1-G^0H^\Delta]^{-1}$ in power series and ignore the higher-order terms. The above equation can then be rewritten as
\begin{equation}
\begin{aligned}
G_\text{out,in}-G^0_\text{out,in} &= G^0_\text{out,in}+[G^0H^\Delta G^0]_\text{out,in} -G^0_\text{out,in}\\
&=[G^0H^\Delta G^0]_\text{out,in}.
\end{aligned}
\end{equation}
It has been established in previous studies that the end-to-end Green's function of a non-Hermitian system with the NHSE can exhibit exponential amplification. We recall that our setup consists of a non-Hermitian lattice system that hosts the NHSE and is operated as a sensor. Suppose our system likewise possesses an exponentially amplified end-to-end Green's function. That is, 
\begin{equation}
G^0_{L,1}\sim e^{\alpha L},
\end{equation}
where we have placed the ``in" and ``out" ports at opposite ends of the system, $L$ is the system size, 
$\alpha$ is a model-dependent constant, and the sign of $\alpha$ depends on the input-output direction. Here, we set the lattice ordering along the skin-effect direction, so that the directional amplification is designed from site $1$ to site $L$. Based on the above discussion, one can get that 
\begin{equation}
\begin{aligned}
G_{L,1}-G^0_{L,1} =& \sum_{\mu\nu} G^0_{L,\mu}H^\Delta_{\mu,\nu} G^0_{\nu, 1}\\
=&G^0_{L,1}H^\Delta_{1,L} G^0_{L,1} + \sum_{\mu=1}^{\mu=L-1}\sum_{\nu=2}^{\nu=L}
 G^0_{L,\mu}H^\Delta_{\mu,\nu} G^0_{\nu, 1}.
\end{aligned}
\end{equation}
Suppose that the second term, i.e., $\sum_{\mu=1}^{\mu=L-1}\sum_{\nu=2}^{\nu=L}
 G^0_{L,\mu}H^\Delta_{\mu,\nu} G^0_{\nu, 1}$, in the above can be neglected. Then the above equation can be written as
\begin{equation}
\begin{aligned}
G_{L,1}-G^0_{L,1} \approx \kappa e^{\alpha L} \Delta,
\end{aligned}
\end{equation}
where $\kappa$ and $\alpha$ are model-specific constants. The last approximation is justified by the fact that the magnitude of the end-to-end Green's function $G^0_{L,1}$ grows exponentially with the distance between the input and output sites. Crucially, the NHSE is essential to this mechanism, as it ensures that the probability amplitude of the wave function accumulates along the skin direction, thereby giving rise to directional amplification of the signal. Moreover, the NHSE is topologically protected by non-Hermitian topology \citep{yao2018edge,yokomizo2019non,okuma2020topological,zhang2020correspondence}; for any one-band model in one dimension, this protection manifests as a nontrivial winding number of the energy spectrum under periodic boundary conditions and the presence of skin modes under open boundary conditions. Consequently, our proposed sensing mechanism inherits this topological protection.

\section*{Theoretical models}
\label{SecIII}
We now instantiate our basic idea through two theoretical models. The first case is a coupled two-chain model, where the coupling originates from the interference of the measurand. The second case can be regarded as a boundary-modification scenario, where the measurand interferes with the coupling between the ends of the open chain.

\subsubsection*{The coupled two-chain model}
We consider a model consisting of two weakly coupled Hatano-Nelson (HN) open chains, which feature distinct asymmetric nearest-neighbor hoppings. The weak coupling, originating from the measurand, mediates the inter-chain interaction. The two chains are configured with opposite NHSE directions. Within this framework, the tight-binding Hamiltonian is given by
\begin{equation}
 	\hat{H}_{coupled} = \hat{H}_A + \hat{H}_B+ \hat{H}_\Delta,
\label{Hamiltonian}
\end{equation}
where
\begin{equation}
\begin{aligned}
	\hat{H}_{A(B)} =& \sum_{n}\left( t_{L(R)}\hat{c}_{n,A(B)}^\dag {\hat{c}_{n+1,A(B)}}+t_{R(L)}{\hat{c}_{n+1,A(B)}}^\dag{\hat{c}_{n,A(B)}}   \right),\\ 
	\hat{H}_\Delta =& \sum_{n}\Delta \hat{c}_{n,A}^\dag\hat{c}_{n,B}.
\end{aligned}
\end{equation}
Here, $c^\dagger_{n,\sigma}(c_{n,\sigma})$ denotes the creation (annihilation) operator at the $n$-th lattice site of chain $\sigma$. The parameters $t_L$ and $t_R$ represent the nearest-neighbor hopping amplitudes, and  $\Delta > 0$ is the interchain coupling strength at each site, with $t_L > t_R > 0$. The exponential amplification of the single-chain Green's function originates from the NHSE, which is driven by the nonreciprocal hopping characterized by $t_L\neq t_R$.

\begin{figure}[htb]
\includegraphics[width=1.0\linewidth]{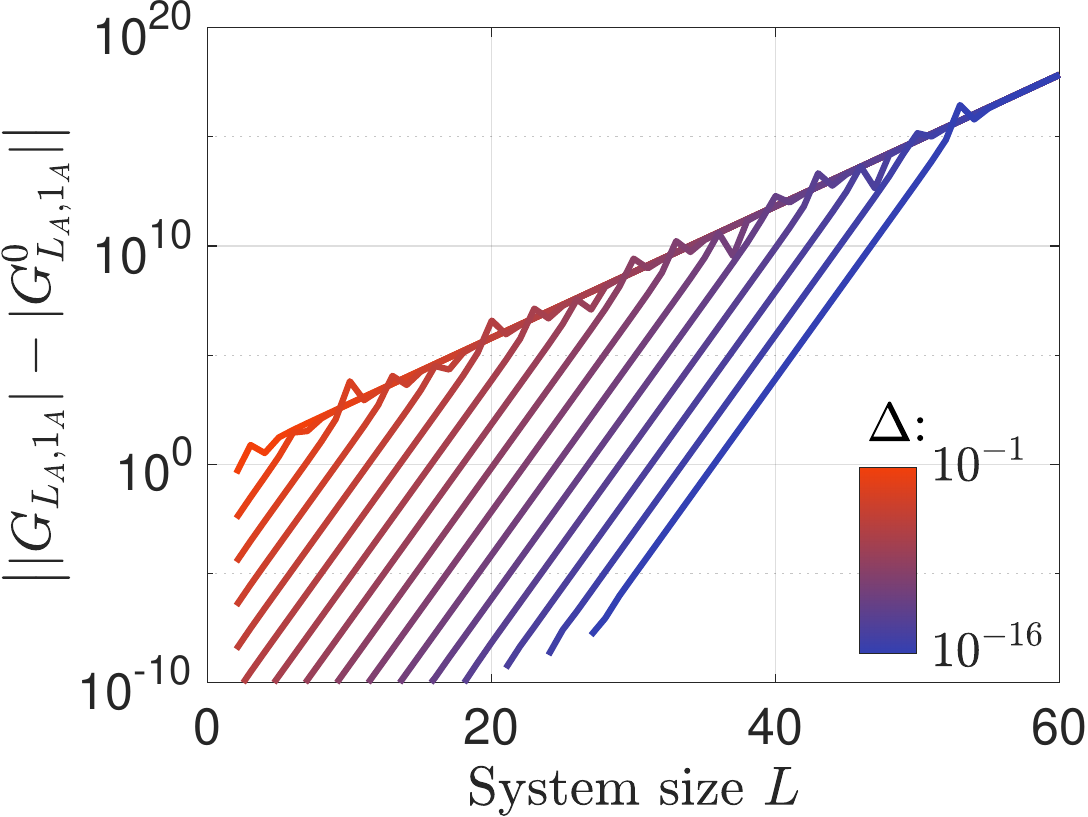}
\caption{Exponential scaling of $\left||G_{L_A,1}|-|G^0_{L_A,1}|\right|$ with the system size $L$ of the coupled two-chain model. The relevant parameters are $t_R = 0.1$, $t_L=1$ and $\omega=0.7$.} 
\label{fig:cNHSE}
\end{figure}

In Fig.~\ref{fig:cNHSE}, we plot the absolute response shift $\left||G_{L,1}(\Delta)|-|G_{L,1}(0)|\right|=\left||G_{L_A,1_A}|-|G^0_{L_A,1_A}|\right|$ induced by a finite interchain coupling $\Delta$, which modifies the Green's function of a single chain. Results are shown for various values of $N$ and $\Delta$. The exact numerical results are found to be in essentially perfect agreement with the approximate analytical predictions, revealing an exponentially amplified signal that spans many orders of magnitude. This quantitatively corroborates our general analysis with high accuracy. We note that the probe frequency $\omega$ is set within the regime $|\omega|<|t_L+t_R|$, which ensures exponentially amplified single-chain Green's functions across all $N$ values considered.

Note that a characteristic system size exists that separates two distinct localization regimes: the skin-localization regime and the scale-free localization (SFL) regime~\citep{li_criticalNHSE_2020,li2021impurity,gou2021exact,yokomizo_scaling_2021,qin_universal_2023,gou2023accumulation,li2023scale,guo2024scale,liang2025intrinsic,ou2025ASL}.
The skin localization is protected by non-Hermitian topology and underpins the exponential amplification mechanism, as reflected in Fig.~\ref{fig:cNHSE}. When the system size exceeds this critical value, most eigenstates enter the SFL regime. In this regime, the eigenstates no longer accumulate at the boundary; instead, they spread over each site $x$ with a characteristic scaling $e^{x/L}$. This implies that as $L$ increases, the eigenstates become increasingly extended, and consequently the exponential amplification of the Green's function ceases to persist.

\subsubsection*{The boundary-modification model}
We consider a Hatano–Nelson (HN) open chain with end-to-end coupling $\Delta$ arising from measurand-induced interference. The corresponding tight-binding Hamiltonian reads
\begin{equation}
\begin{aligned}
{\hat{H}_{boundary}} =& \hat{H}^0 + \hat{H}^\Delta\\
=&\sum_n\left(t_L\hat{c}_{n}^\dag {\hat{c}_{n+1}}+t_R{\hat{c}_{n+1}}^\dag{\hat{c}_{n}} \right) + \Delta\left(\hat{c}_{L}^\dag {\hat{c}_{1}}+{\hat{c}_{1}}^\dag{\hat{c}_{L}}\right).
\end{aligned}
\end{equation}
Here,  $c^\dagger_{n}(c_{n})$ denotes the creation (annihilation) operator at the $n$-th lattice site. The parameters $t_L$ and $t_R$ are the nearest-neighbor hopping amplitudes, and $\Delta > 0$ is the end-to-end coupling strength, with $t_L > t_R > 0$. The system is assumed to contain an even number of sites, specifically $2L$. 
\begin{figure}[htb]
\includegraphics[width=1.0\linewidth]{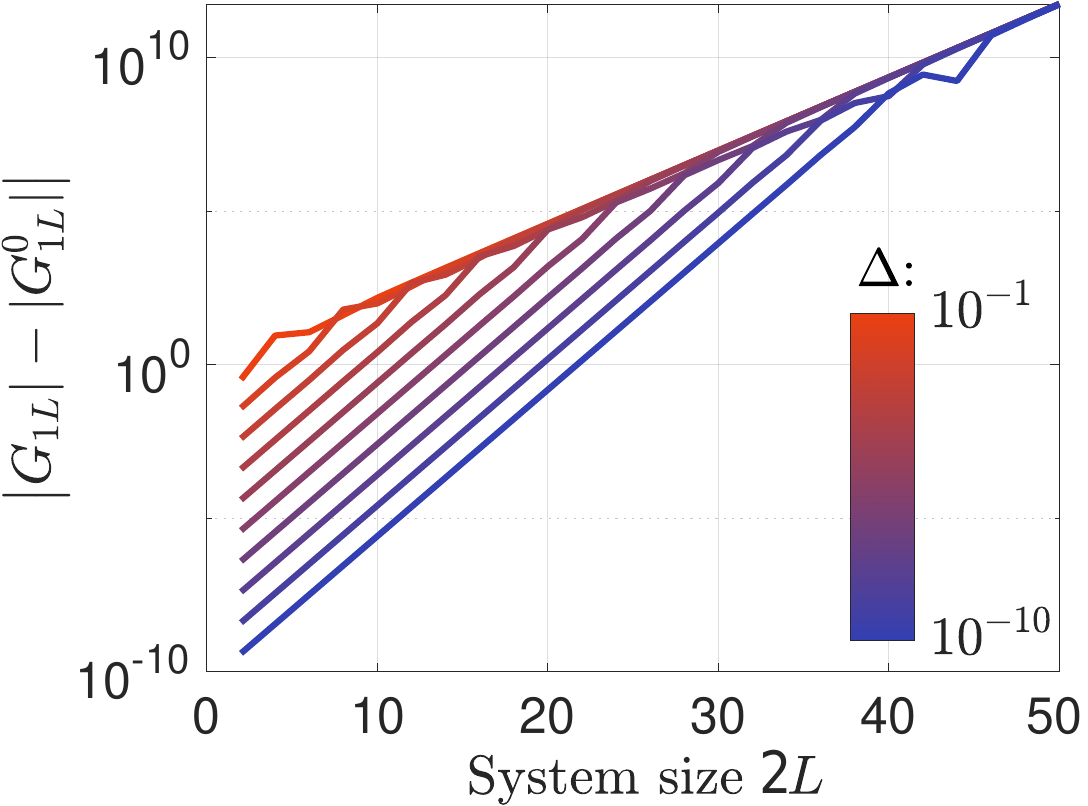}
\caption{Exponential scaling of the response discrepancy $\left||G_{L,1}|-|G^0_{L,1}|\right|$ versus the system size $2L$ in the one-band-chain-based non-Hermitian topological sensor. Parameters: $t_R=1.5$, $t_L=0.5$ and $\omega = 0$.} 
\label{fig:HN}
\end{figure}

Fig.~\ref{fig:HN} reveals that the absolute response shift $\left||G_{L,1}(\Delta)|-|G_{L,1}(0)|\right|=\left||G_{L,1}|-|G^0_{L,1}|\right|$ exhibits the same behavior as in the coupled two-chain model. This plot demonstrates that the exponential amplification mechanism breaks down for sufficiently large system sizes, a result consistent with Fig.~\ref{fig:cNHSE}, indicating that the underlying physics is the same in both models. For sufficiently small systems, the perturbation-induced reconstruction of the spectrum and eigenstates remains weak, and the system retains its skin-localized character, thereby sustaining the exponential Green’s-function amplification. As the system size increases, the enhanced non-normal sensitivity makes the weak perturbation increasingly effective, eventually driving the system into the SFL regime~\citep{trefethen2020spectra,nakai2024topological,liang2025intrinsic}. In this regime, as the system size continues to increase, the energy spectrum increasingly approaches the PBC spectrum, and the eigenstates become progressively more extended.

We emphasize that the numerical results in Fig.~\ref{fig:HN} are restricted to the even-lattice configuration. The odd-lattice counterpart, however, is susceptible to numerical instabilities that can contaminate the data. To rule out such artifacts and ensure the integrity of our analysis, we present an exact analytical derivation for the odd-lattice case in the Appendix.

\section*{Robustness and non-Hermitian topology}
\label{SecIV}
A realistic system inevitably harbors some degree of disorder. Here, we turn to the question of how the non-Hermitian topology protects our scheme against such imperfections. To this end, we introduce disorder into the boundary-modification model (with all other parameters identical to those in Fig.~\ref{fig:HN}), given by
\begin{equation}
\hat{H}_\text{disorder} = \sum_{n=1}^{L-1} \left(W_+(n)\hat{c}_{n}^\dagger\hat{c}_{n+1} + W_-(n)\hat{c}_{n+1}^\dagger\hat{c}_{n}\right).
\end{equation}
Here, $W_+(n)$ and $W_-(n)$ are random variables uniformly drawn from the interval $[-W,W]$, where $W>0$ denotes the disorder strength. For a wide range of parameters, the system retains a nontrivial spectral winding number $w(E_r)$  as long as $W$ remains sufficiently small. In the presence of disorder, the spectral winding number $w(E_r)$ is defined as follows ~\citep{claes2021skin}
\begin{equation}
w(E_r) = \mathcal{T}(\hat{Q}^\dagger [\hat{Q},\hat{X}]),
\end{equation}
where $\mathcal{T}$ denotes the trace per unit volume, $\hat{Q}$ is the unitary operator obtained from the polar decomposition $(\hat{H}-E_r)=\hat{Q}\hat{P}$, and $\hat{X}$ is the position operator. Here, $E_r$ is the reference energy. 

\begin{figure}[htb]
\includegraphics[width=1.0\linewidth]{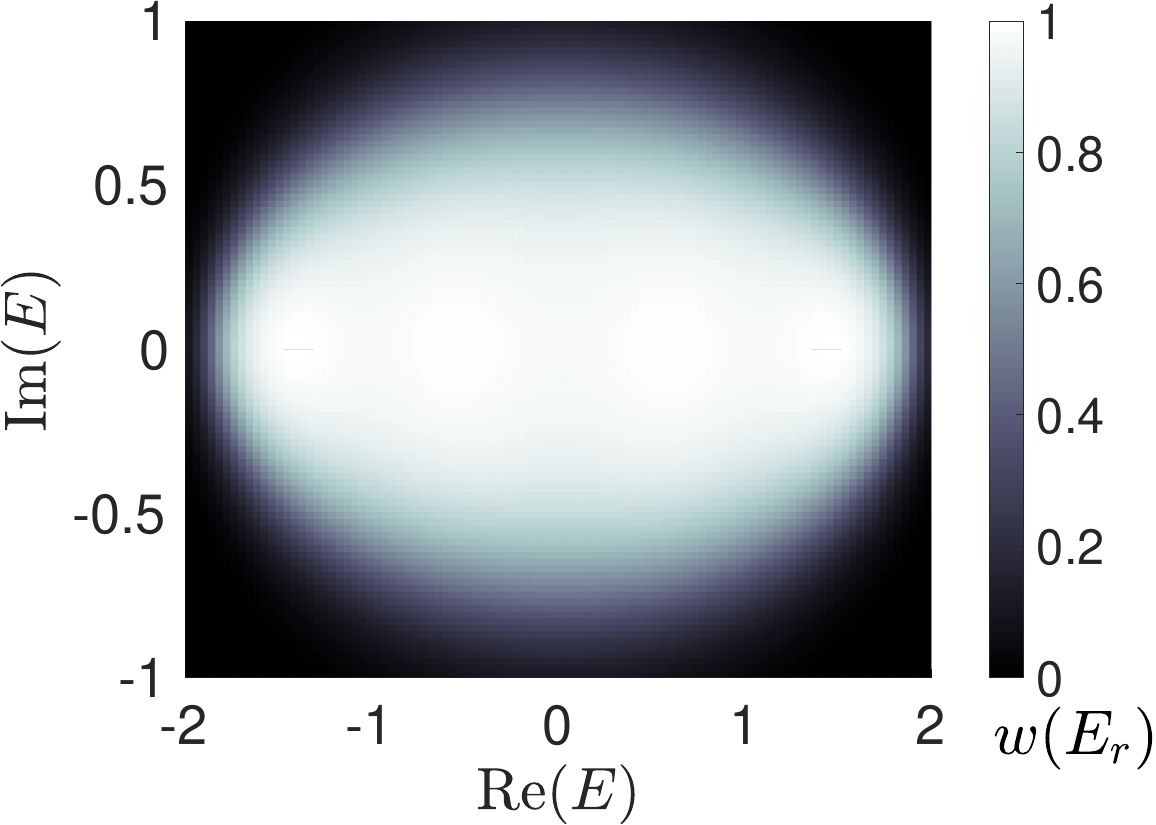}
\caption{The disordered spectrum winding number for chain A. The parameters are $\Delta=10^{-7}$, $L=16$. All other parameters are same as in Fig.~\ref{fig:HN}.}
\label{fig:disorderWinding}
\end{figure}

\begin{figure}[htb]
\includegraphics[width=1.0\linewidth]{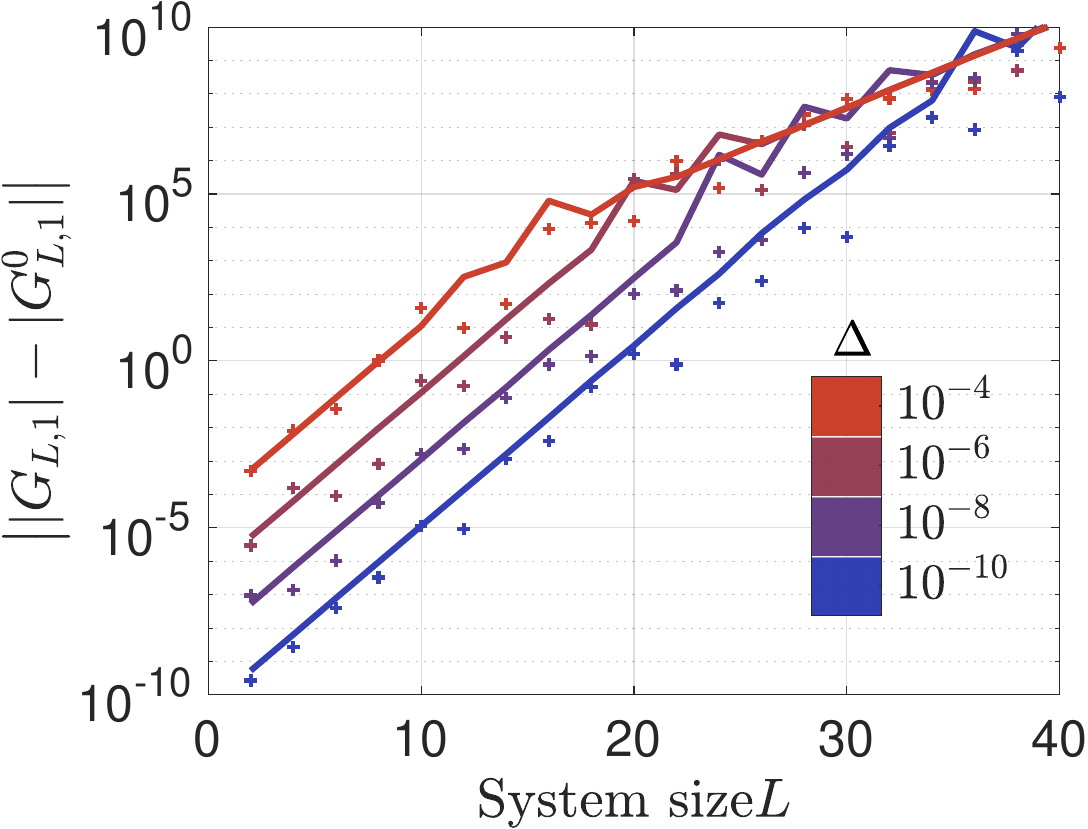}
\caption{Scaling of the Green's function $\left||G_{L,1}|-|G_{L,1}^0|\right|$ of one band case, for various $\Delta$, with disorder strength $W = 0.2$. The input and output are fixed to chain $A$. Data for a single disorder realization (marked by "+") are compared with the disorder-averaged result (line) over $M=10^4$ realizations for each $L$. All other parameter are same as in Fig.~\ref{fig:HN}.}
\label{fig:HNdisorder}
\end{figure}

The exponential amplification mechanism originates from the NHSE, which is protected by a nontrivial spectral winding number. In Fig.~\ref{fig:disorderWinding}, we plot the disorder-resolved spectral winding number for chain A.  A value of $w(E_r) = +1$ indicates that the skin modes accumulate from site 1 to site $L$. Conversely, $w(E_r) = 0$ signals the disappearance of the skin effect, while $w(E_r) = -1$ corresponds to reversed accumulation, causing the amplification mechanism to fail and transition into exponential decay. In the latter case, however, the exponential amplification can be restored by reversing the roles of the ``in" and ``out" channels of the Green's function. This implies that even when disorder flips the sign of the nontrivial spectral winding number, our theory remains applicable—one simply needs to interchange the ``in" and ``out" designations. This result is corroborated and exemplified in Fig.\ref{fig:HNdisorder}:  for individual disorder realizations, the exponential growth of the sensitivity $\mathcal{S}$ is clearly visible, albeit with noticeable sample-to-sample fluctuations. These fluctuations are eliminated upon averaging over a sufficiently large number of disorder realizations for each system size $L$.

\section*{Experimental platforms}
\label{SecV}
To realize our setup experimentally, we propose to adopt the driven dissipative cavity-array platform described in Refs.~\citep{wanjura2020topological,liang_anomalous_2022}, with two arrays coupled coherently. The Hamiltonian and the master equation governing this proposed experimental system are given by
\begin{equation}
\begin{aligned}
\mathcal{H} &= \sum_n\left(J_A \hat{c}_{n,A}^\dagger \hat{c}_{n+1,A} + J_b \hat{c}_{n,B}^\dagger \hat{c}_{n+1,B} + J_0 \hat{c}_{n,A}^\dagger \hat{c}_{n,B}+\text{H.c.}\right),\\
\dot{\rho} &= -i[\mathcal{H},\rho] + \sum_n\left(\Gamma_\sigma \mathcal{D}[\hat{z}_{n,\sigma}]\rho+\gamma_\sigma\mathcal{D}[\hat{a}_{n,\sigma}]\rho
+\epsilon\mathcal{D}[\hat{a}^\dagger_{n,\sigma}]\rho\right).
\end{aligned}
\end{equation}
Here, $\mathcal{D}[\hat{z}_{n,\sigma}]\rho = \hat{z}_{n,\sigma}\rho\hat{z}^\dagger_{n,\sigma} -\frac{1}{2}\{\hat{z}_{n,\sigma}^\dagger\hat{z}_{n,\sigma},\rho\}$ with $\hat{z}_{n,\sigma} = \hat{c}_{n,\sigma} + e^{-i\theta_\sigma}\hat{c}_{n,\sigma}$  (assuming nearest-neighbor coupling). The parameters $J_\sigma$ and $J_0$ denote the intrachain and interchain hopping amplitudes, respectively, with $\sigma = A$ or $B$. The term $\Gamma_\sigma \mathcal{D}[\hat{z}_{n,\sigma}]\rho$ accounts for dissipative nearest-neighbor couplings; $\epsilon\mathcal{D}[\hat{a}^\dagger_{n,\sigma}]\rho$ describes uniform local incoherent photon pumping, which drives the system into a regime characterized by a nontrivial spectral winding number; and $\gamma_\sigma\mathcal{D}[\hat{a}_{n,\sigma}]\rho$ represents photon decay into a waveguide coupled to each cavity, which simultaneously provides input and output ports for the signal.

For an input field $\langle c_{n,\sigma}^\text{in}(t)\rangle$ injected into the system via the waveguide, the equations of motion for the mean cavity-field amplitudes $\langle c_{n,\sigma}\rangle$ take the form~\citep{wanjura2020topological,wanjura2021correspondence}
\begin{equation}
\begin{aligned}
\langle \dot{c}_{n,\sigma}\rangle =& \left(\frac{\epsilon-\gamma_\sigma-2\Gamma_\sigma}{2}\right)\langle c_{n,\sigma}\rangle - i\Delta\langle c_{n,\sigma}\rangle - \sqrt{\gamma_\sigma}\langle c_{n,\sigma}^\text{in}\rangle\\
&+ \left(iJ_\sigma + \frac{e^{-i\theta_\sigma}\Gamma_\sigma}{2}\right)\langle c_{n+1,\sigma}\rangle
+\left(iJ_\sigma + \frac{e^{i\theta_\sigma}\Gamma_\sigma}{2}\right)\langle c_{n-1,\sigma}\rangle\\
\equiv& -i \sum_{n^\prime}\sum_{\sigma^\prime = A,B} H_{n\sigma,n^\prime\sigma^\prime}\langle c_{n^\prime,\sigma^\prime}\rangle - \sqrt{\gamma_\sigma}\langle c_{n,\sigma}^\text{in}\rangle.
\end{aligned}
\label{eq:waveguide}
\end{equation}
where $\sigma\neq \sigma^\prime$. The input and output fields, $\langle c_{n,\sigma}^\text{in}(\omega)\rangle$ and $\langle c_{n,\sigma}^\text{out}(\omega)\rangle$, at frequency $\omega$ are related via the input-output formalism as~\citep{wanjura2020topological,wanjura2021correspondence}
\begin{equation}
\textbf{a}_\text{out} = \textbf{a}_\text{in} - \frac{i\gamma}{\omega-H}\textbf{a}_\text{in} = \textbf{a}_\text{in} - i\gamma G\textbf{a}_\text{in},
\label{eq:G}
\end{equation}
with $\textbf{a}_\text{out/in} = \left(\langle c_{1,A}^\text{out/in}\rangle,\cdots,\langle c_{L,A}^\text{out/in}\rangle,\langle c_{1,B}^\text{out/in}\rangle,\cdots,\langle c_{L,B}^\text{out/in}\rangle\right)^T$. Note that the Green’s function $G$ in Eq.~\eqref{eq:G} is defined with a real frequency $\omega$, which corresponds to the input signal frequency in the preceding discussion. Its imaginary part can be effectively introduced via a uniform local pumping of strength $\epsilon$, which shifts the spectrum along the imaginary axis.

\begin{figure}
\includegraphics[width=1.0\linewidth]{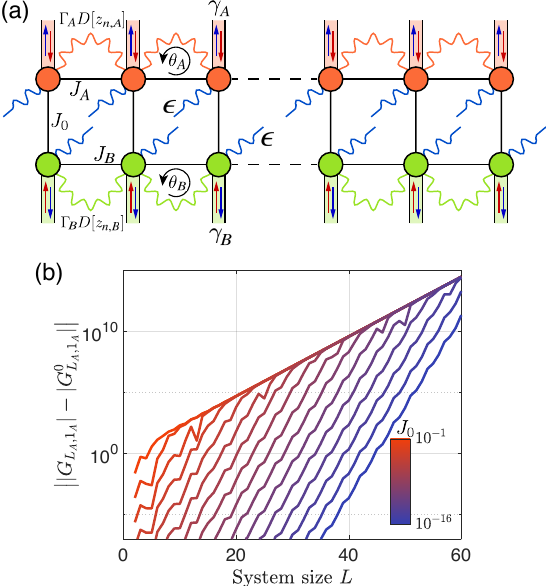}
\caption{(a) Driven-dissipative coupled chains consisting of 2N bosonic cavity modes, described by Eqs.~\eqref{eq:waveguide}. (b) Numerically simulated results for the proposed setup. The relevant parameters are $J=0.75$, $\theta = 0.75\pi$, $\Gamma=0.5$, $\epsilon=2$, and $\gamma_A = \gamma_B = 1$.} 
\label{fig:waveguide}
\end{figure}

We now turn to the steady-state response of the proposed system, considering a signal that enters and exits at distinct ends. Specifically, a signal injected at $(1,A)$ and detected at $(L-1,A)$ is described by
\begin{equation}
\langle c_{L,A}^\text{out}\rangle = -i\gamma G_{(L-1)_A,1_A}\langle c_{1,A}^\text{in}\rangle,
\end{equation}
with $G_{Na,1a}$ an element of the Green’s function. The effective Hamiltonian obtained from the model in Eq.~(\ref{Hamiltonian}) can then be mapped onto the proposed experimental setup under the following conditions
\begin{equation}
t_L = J-i\frac{e^\theta \Gamma}{2},\quad t_R = J-i\frac{e^{-\theta} \Gamma}{2}\quad \Delta=J_0,
\end{equation}
where $\theta_A = -\theta_B=\theta$, $J_A=J_B=J$, $\epsilon-\gamma_\sigma-2\Gamma_\sigma=0$ and $\Gamma_A=\Gamma_B=\Gamma$. In Fig.~\ref{fig:waveguide}(b), we display the scaling of the Green's function discrepancy $||G_{L_A,1_A}|-||G^0_{L_A,1_A}||$ for the proposed implementation. The setup achieves superior performance for smaller measurand values.

\section*{Conclusion}\label{SecVI}
We have proposed a non-Hermitian sensing scheme that achieves exponential sensitivity scaling via the end-to-end Green's function amplification arising from the NHSE. This mechanism is validated through two distinct models: a coupled two-chain system and a boundary-modification model. In both cases, the exponential enhancement persists up to a critical system size, beyond which the system enters the scale-free localization regime and the amplification ceases. The scheme is topologically protected by a nontrivial spectral winding number, ensuring robustness against moderate disorder. When disorder reverses the winding number sign, the amplification can be restored by interchanging the input and output ports. Numerical simulations confirm these predictions. We further demonstrate feasibility in driven-dissipative cavity arrays via input-output theory, providing a concrete path toward experimental realization. Our work establishes a framework linking non-Hermitian spectral topology and directional response amplification to enhanced sensing, offering both high sensitivity and robustness against disorder.

\textit{\it \clr Acknowledgments.} The authors acknowledge support from Shandong Provincial Natural Science Foundation of China through Grant No. ZR2023MA048.


\section*{Appendix}
\subsection*{Green’s-function amplification in the coupled two-chain model}\label{AppxA}

We now derive the end-to-end Green's function for the coupled two-chain model to explicitly demonstrate the exponential amplification mechanism. The frequency-domain Green's function is defined as $G=(\omega-H)^{-1}$. The 
$(x_\text{out}=L,x_\text{in}=1)$ entry, which corresponds to the amplification direction from site 1 to site 
$L$, is given by
\begin{equation}
\begin{aligned}
G_{L_A,1_A} &= (\omega -H)^{-1}_{L_A,1_A}  \\
&=\frac{\text{adj}(\omega-H)_{L_A,1_A}}{\text{Det}[\omega - H]}\\
&=\frac{(-1)^{L_A+1_A}\text{Det}[A^\prime]\text{Det}[B-C_1(A^\prime)^{-1}C_2]}{\text{Det}[A]\text{Det}[B-CA^{-1}C]},
\end{aligned}
\end{equation}
where 
\begin{eqnarray*}
A^\prime &= \begin{pmatrix}
-t_L	&0	&\cdots	&0 &0\\
\omega &-t_L	&\cdots	 &0 &0\\
\vdots	&\vdots	&\ddots	&\vdots &\vdots \\
0	&0	&\cdots	&-t_L	&0\\
0	&0	&\cdots	&\omega	&-t_L\\
\end{pmatrix}_{(L-1)\times(L-1)},\\
C_1 &= \begin{pmatrix}
0	&0	&\cdots	&0\\
-\Delta &0	&\cdots	&0\\
0 	&-\Delta	&\cdots	&0\\
\vdots	&\vdots	&\ddots	&\vdots\\
0	&\cdots	&0 	&-\Delta\\
\end{pmatrix}_{(L-1)\times(L)},\\
C_2 &= \begin{pmatrix}
&-\Delta &0	&\cdots	&0	&0\\
&0 	&-\Delta	&\cdots	&0	&0\\
&\vdots	&\vdots	&\ddots	&\vdots	&\vdots\\
&0	&\cdots	&0 	&-\Delta	&0\\
\end{pmatrix}_{(L)\times(L-1)},\\
\end{eqnarray*}
and $A = \omega-H_A$, $B = \omega-H_B$, $C = -H_\Delta$. Note that the free Green's function of chain A constitutes the unperturbed part of $G=(\omega-H)^{-1}$, allowing us to rewrite the full Green's function as
\begin{equation}
\begin{aligned}
&\frac{(-1)^{L_A+1_A}\text{Det}[A^\prime]\text{Det}[B-C_1(A^\prime)^{-1}C_2]}{\text{Det}[A]\text{Det}[B-CA^{-1}C]} \\
=& G^0_{L_A,1_A}\frac{\text{Det}[B-C_1(A^\prime)^{-1}C_2]}{\text{Det}[B-CA^{-1}C]}\\
=& G^0_{L_A,1_A} M,
\end{aligned}
\end{equation}
with a modified factor $M$. 

We first compute the Green's function of chain A, denoted as $G^0_{L_A,1_A}$. The determinant Det$[A]$ can be reformulated using the transfer matrix form. Specifically, the iterative equations
\begin{equation}
D_L = \omega D_{L-1} + (-1)(-t_L)(-t_R)D_{L-2},
\end{equation}
with denoting the determinant Det$[A]$ by
\begin{equation}
D_L :=
\left\vert
\begin{matrix}
\omega  	&-t_L		&0 		&\cdots		&0	&0\\
-t_R		&\omega 	&-t_L 	&\cdots		&0	&0\\
0		&-t_R		&\omega  	&\cdots		&0	&0\\
\vdots	&\vdots	&\vdots	&\ddots	&\vdots	&\vdots\\
0		&0		&0		&\cdots		&\omega  	&-t_L\\\
0		&0		&0		&\cdots		&-t_R		&\omega\\
\end{matrix}
\right\vert_{L\times L},
\end{equation}
can be rewritten as
\begin{equation}
\begin{aligned}
\begin{pmatrix}
D_L\\
D_{L-1}
\end{pmatrix}
&=
\begin{pmatrix}
\omega		&-t_Lt_R\\
1		&0\\
\end{pmatrix}
\begin{pmatrix}
D_{L-1}\\
D_{L-2}\\
\end{pmatrix}\\
&=
\begin{pmatrix}
\omega		&-t_Lt_R\\
1		&0\\
\end{pmatrix}
\begin{pmatrix}
\omega		&-t_Lt_R\\
1		&0\\
\end{pmatrix}
\cdots
\begin{pmatrix}
D_2\\
D_1\\
\end{pmatrix}\\
&=
\begin{pmatrix}
\omega		&-t_Lt_R\\
1		&0\\
\end{pmatrix}
\begin{pmatrix}
\omega		&-t_Lt_R\\
1		&0\\
\end{pmatrix}
\cdots
\begin{pmatrix}
\omega		&-t_Lt_R\\
1		&0\\
\end{pmatrix}
\begin{pmatrix}
\omega\\
1\\
\end{pmatrix}\\
&=[T(\omega)]^{L-1}
\begin{pmatrix}
\omega\\
1\\
\end{pmatrix},\\
\end{aligned}
\end{equation}
with the initial conditions $D_1= \omega$ and $D_2=\omega D_1-t_Lt_R$. This second-order matrix difference equation can be solved straightforwardly. 
The characteristic polynomial of the transfer matrix $T(\omega)$ is $z^2=z\omega-t_Lt_R$, whose two roots are
 \begin{equation}
 z_\pm = \frac{\omega\pm\sqrt{\omega^2-4t_Lt_R}}{2}.
 \end{equation}
The determinant can therefore be written as
\begin{equation}
D_L = \frac{z_+^{L+1}-z_-^{L+1}}{z_+-z_-}.
\end{equation}

With the above results, one can get that
\begin{equation}
\begin{aligned}
G^0_{L_A,1_A} &= \frac{\text{adj}(\omega-H_A)_{L,1}}{D_L} = \frac{t_L^{L-1}}{D_L}\\
&= t_L^{L-1}\frac{z_+-z_-}{z_+^{L+1}-z_-^{L+1}},
\end{aligned}
\end{equation}
where $\beta_\pm = z_\pm/t_L$. Consider the case $|\beta_+|>|\beta_-|$ for a large system size $L\gg 1$. The Green's function $G^0_{1_A,L_A}$ then satisfies the scaling $|\beta_+|^{-(L-1)}$, that is,
\begin{equation}
G^0(\omega)_{L_A,1_A}\sim\left\lbrace
\begin{aligned}
L^0		\qquad &|\beta_-|\approx|\beta_+|,\ 0<\omega = t_L+t_R,\\
e^{\alpha L}	\qquad &|\beta_-|<|\beta_+|<1,\ 0<\omega < t_L+t_R,\\
e^{-\alpha L} \qquad &1<|\beta_-|<|\beta_+|,\ 0 < t_L+t_R<\omega.\\
\end{aligned}\right.
\label{eq:freeG}
\end{equation}

We now present the perturbative solutions for $M$ up to second order in the coupling parameter 
$\Delta$. They can be explicitly written as
\begin{equation}
M = (1 + P \Delta^2)^{-1}.
\end{equation}
The Green's function $G(\omega)_{1_A,L_A}$ reduces to $G_A(\omega)_{1,L}$ when $\Delta = 0$ or $P \ll 1/\Delta^2$. In the main text, we have chosen $t_L>t_R$ so that the amplification is directed leftward. Without loss of generality and for convenience, we now consider the limiting case $t_R \ll t_L,\omega$. In this limit, the polynomial $P$ can be explicitly written as
\begin{equation}
\begin{aligned}
P|_{t_R\rightarrow 0} &= (-1)\frac{(L-1)\omega^{2(L-1)}+\sum_{j=1}^L j t_L^{2(L-j)+1}\omega^{2(j-1)}}{t_L\omega^{2L}}\\
&=(-1)\frac{(L-1)\omega^{2(L-1)}+\frac{t_L^{3+2L}}{(t_L^2-\omega^2)^2}}{t_L\omega^{2L}}\\
&=\frac{1-L}{t_L \omega^2}-\frac{t_L^{2 L+2} \omega^{-2 L}}{\left(t_L^2-\omega^2\right)^2}.\\
\end{aligned}
\end{equation}
For a large size $L > L_c \gg 1$, the scaling of modified factor $|M|$ satisfies
\begin{equation}
\left|M\right|\sim\left\lbrace
\begin{aligned}
|\kappa|^{-1} e^{-\alpha^\prime L}\Delta^{-2}\qquad &\omega < t_L,\\
|c|^{-1}L^{-1}\Delta^{-2}	\qquad 	&\omega \geq t_L.\\
\end{aligned}\right.
\end{equation}
For $1 < L < L_c$, the scaling satisfies $L^0$ because $P \ll 1/\Delta^2$. We expect that the scaling of $G_{L_A,1_A}$ obeys the same conditions for $t_L > t_R$. The complete scaling behavior of $G_{L_A,1_A}$ is summarized in Table \ref{scaling_G1L}. 

\begin{table}[htb]
\begin{center}
\caption{The Scaling of $G_{L_A,1_A}$}
\label{scaling_G1L}
\begin{tabular}{|c|c|c|}
\hline
$L>L_c$	&	$L<L_c$	&	\\
\hline
$e^{(\alpha-\alpha^\prime) L}\Delta^{-2}$	&$e^{\alpha L}$ 	& $|\omega| < |t_L+t_R|$\\
\hline
$L^0$	&$L^0$ 	& $|\omega| = |t_L+t_R|$\\
\hline
$e^{-\alpha L}$	&$e^{-\alpha L}$ 	& $|\omega| > |t_L+t_R|$\\
\hline
\end{tabular}
\end{center}
\end{table}

\begin{figure}[htb]
\includegraphics[width=1.0\linewidth]{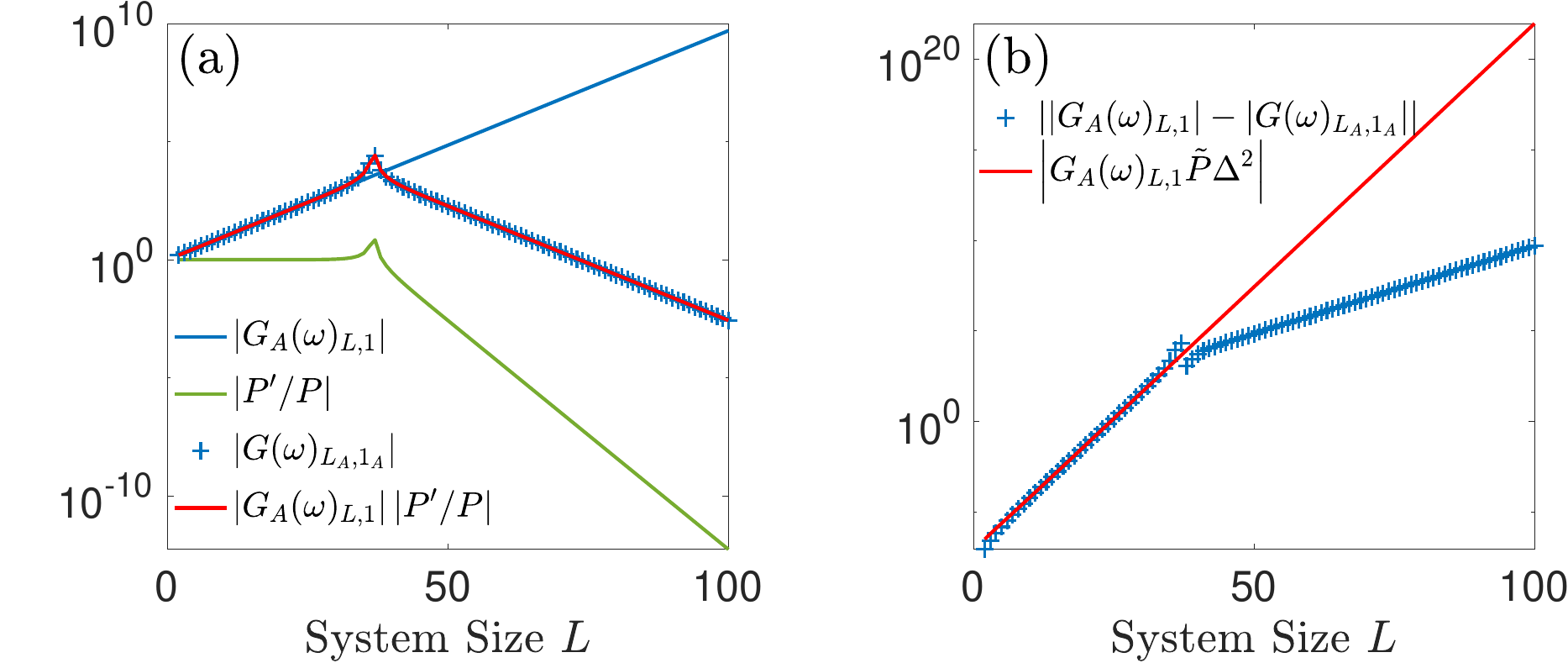}
\caption{End-to-end Green's function for the one-band case, with input and output fixed to chain $A$. (a) The symbol '+' represents the numerical results. The blue, red, and green lines respectively mark the decoupled case (numerical), the decoupled case with the modified factor $|P^\prime/P|$, and the modified factor $|P^\prime/P|$ itsefly. (b) The analytical results are fitted to the numerical data in the regime $L<L_c$. The parameters are $\omega = 0.8$, $t_L = 1$, $\Delta = 10^{-4}$ and $t_R=0.0001$.} 
\end{figure}

\subsection*{Green’s-function amplification in the boundary-modification model}\label{AppxB}

The Green's function of the model reads
\begin{equation}
G = G^0 + G^0TG^0,
\end{equation}
where $T$-matrix is 
\begin{equation}
\begin{aligned}
T =& f\left(\frac{\ket{L}\Delta\bra{1}}{1-\Delta\bra{L}G^0\ket{1}}+\frac{\ket{1}\Delta\bra{L}}{1-\Delta\bra{1}G^0\ket{L}}\right.\\
&\left. + \frac{\ket{L}\Delta\bra{1}}{1-\Delta\bra{L}G^0\ket{1}}G^0\frac{\ket{1}\Delta\bra{L}}{1-\Delta\bra{1}G^0\ket{L}}\right.\\
&\left.+ \frac{\ket{1}\Delta\bra{L}}{1-\Delta\bra{1}G^0\ket{L}}G^0\frac{\ket{L}\Delta\bra{1}}{1-\Delta\bra{L}G^0\ket{1}}\right),
\end{aligned}\label{eq:T}
\end{equation}
and the Green's function $G^0$ is $[\omega-H^0]^{-1}$, with
\begin{equation}
f = \frac{1}{1-\frac{\Delta}{1-\Delta\bra{L}G_0\ket{1}}\frac{\Delta}
{1-\Delta\bra{1}G_0\ket{L}}\bra{1}G_0\ket{1}\bra{L}G_0\ket{L}}.
\end{equation}
The difference between the end-to-end green's functions $G$ and $G^0$ is
\begin{equation}
[G-G^0]_{L,1} = [G^0TG^0]_{L,1}.
\label{eq:GG0}
\end{equation}
According to the property of the Green's function $G^0$ and for the system size $L\gg 1$, Eq.~\eqref{eq:T} reduces to
\begin{equation}
\begin{aligned}
T =& \frac{1}{1-\frac{\Delta^2\omega^{-2}}{1-\Delta\bra{L}G_0\ket{1}}}
\left(\frac{\ket{L}\Delta\bra{1}}{1-\Delta\bra{L}G^0\ket{1}}+\ket{1}\Delta\bra{L}  \right.\\
&+\left.\frac{\ket{L}\Delta^2\omega^{-1}\bra{L}}{1-\Delta\bra{L}G^0\ket{1}}+\frac{\ket{1}\Delta^2\omega^{-1}\bra{1}}{1-\Delta\bra{L}G^0\ket{1}}\right)\\
&= T_{L1}+T_{1L}+T_{LL}+T_{11},\\
\end{aligned}
\end{equation}
where $T_{ij} = \bra{i}T\ket{j}$.
Substituting Eq.~\eqref{eq:T} into Eq.~\eqref{eq:GG0}, we obtain
\begin{equation}
[G-G^0]_{L,1} = G^0_{LL}T_{L1}G^0_{11}+G^0_{L1}T_{1L}G^0_{L1}+G^0_{LL}T_{LL}G^0_{L1}
+G^0_{L1}T_{11}G^0_{11}.
\end{equation}
The values of the Green's functions $G^0_{LL}$ and $G^0_{11}$ are both $\omega^{-1}$. The value of the Green's function $G^0_{L1}$ is the same as that in Eq.~\eqref{eq:freeG}. Substituting the values of these Green's functions into the above equation, and taking the conditions $0<\omega<t_L+t_R$ and $L\gg 1$, we obtain 
\begin{equation}
[G-G^0]_{L,1} \approx T_{1L}\left(G^0_{L1}\right)^2 \sim \kappa e^{\alpha L}\Delta,
\end{equation}
where $\kappa$ and $\alpha>0$ are the system parameters.

\bibliography{sensing}

\end{document}